# IDENTIFYING DATA AND INFORMATION STREAMS IN CYBERSPACE: A MULTI-DIMENSIONAL PERSPECTIVE

Ikwu Ruth[1] and Louvieris Panos[2]

[1]Department of Computer Sciences, Brunel University, London
ruth.ikwu@brunel.ac.uk

[2][1]Department of Computer Sciences, Brunel University, London
panos.louvieris@brunel.ac.uk

## ABSTRACT

*Cyberspace has gradually replaced the physical reality, its role evolving from a simple enabler of daily live processes to a necessity for modern existence. As a result of this convergence of physical and virtual realities, for all processes being critically dependent on networked communications, information representative of our physical, logical and social thoughts are constantly being generated in cyberspace. The interconnection and integration of links between our physical and virtual realities create a new hyperspace as a source of data and information. Additionally, significant studies in cyber analysis have predominantly revolved around a single linear analysis of information from a single source of evidence (The Network). These studies are limited in their ability to understand the dynamics of relationships across the multiple dimensions of cyberspace. This paper introduces a multi-dimensional perspective for data identification in cyberspace. It provides critical discussions for identifying entangled relationships amongst entities across cyberspace.*

## 1. INTRODUCTION

Information in cyberspace forms a new thinking space for the human and physical components that integrates computing, networks and physical processes. Cyberspace has, therefore, been defined by experts [1], [2] as an artificial world where humans navigate in an information-based space. There is no standard definition of the term 'cyber', but the term is mostly used as a prefix for ideas related to internet networks, information technology and virtual realities. Information security experts [2], [3] mostly view the cyberspace as a network of internet and digital entities that support information flow. An interesting term cited in [4] refers to cyberspace as a virtual and physical abstraction with all the information resources that create the interconnections among cyber entities. As a dynamic, interdependent environment, which is geographically less constrained than other environments, access is possible through virtual applications running on digital devices that store, process, and transmit data [5].

Over the last ten years, cyberspace has gradually replaced the physical reality, and these digital communication devices have been deeply embedded in human lives [6]. The physical reality is the state of things as they exist in the real world, while virtual reality refers to the state of things as they exist in the virtual information space. Therefore, the role of cyberspace has progressively evolved from being a simple enabler of daily activities to a critical necessity for modern existence. As technology evolves, so does the dependence of modern society on networked infrastructure for communication and connectivity. Shopping, financial transactions, communication, crime, education, entertainment are traditional processes of everyday life that have been replaced by digital processes in cyberspace. Cyberspace is, therefore, the link or gateway between the physical and virtual reality-a simple collection of multiple computing devices connected by networks for storing, transmitting and utilising electronic information. An alternative way to understand the nature of cyberspace is to demystify its purpose or the role it plays in human existence.

- *Cyberspace as a medium for inter-connectivity (The physical Space)*

The physical infrastructure and logical building blocks of cyberspace support capabilities for connectivity between linked devices. The rules and nature of operations in cyberspace are defined on the logical layer. For the physical infrastructure, it is important to note the main components that support connectivity: sensors, actuators and networks. Sensors capture physical data and are low-cost devices with limited storage and computational capacities. Actuators, automate, control and convert the collected physical data into action commands. Intelligent computer networks support the linking of digital devices and transmission of physical data through cyberspace. The physical objects establish connections and relations with other cyber entities in both real and digital worlds based on rules defined on the logical layer. Moreover, physical objects are owned and operated by human beings and therefore reflect social attributes that are directly or indirectly correlated with the affiliated persona.

- *Cyberspace as a medium for inter-communication (The social Space):*

Real-world personas control cyber-personas in cyberspace. The persona and cyber-persona are sometimes referred to as the offline and online modes, respectively, of human beings in cyberspace [6]. The logical infrastructure of cyberspace provides capabilities for interaction and communication between multiple cyber-personas, which in turn creates greater capabilities for sharing and collaboration in cyberspace. Sharing refers to an integration of social attributes and social events, exchange of thoughts and the creation of social inter-/intra relationships owned by real-world personas. Collaboration refers to a coming together of cyber-personas in real-time, mainly to create intellectual content. This capability relies on communication infrastructure to support human learning and co-existence. Therefore, cyberspace as a medium for communication creates a social space of intelligent networks with streams of thoughts, perceptions, social affiliation and social relationships.

- *Cyberspace as an Enabler of beliefs (SPEC Factors):*

Given the social capabilities provided by the infrastructure of cyberspace, over the past three decades, cyberspace has also gradually become an influencer of beliefs. Real-world persons are affiliated with various social, political, economic and cultural ideas which shape ideologies. Cyberspace provides a medium for sharing ideologies and creating online communities based on ideological differences. Agents of ideological formation embedded in online social communities propagate ideas, notions and positions that support certain ideological perspectives. Finally, Ning et al. [4] summarise these functions of cyberspace in the formation of a cyber-human hyper-thinking space consisting of data, information, knowledge and thoughts. Data collected by sensors, being exchanged during communication sessions from ubiquitous devices are expected to create context-aware information that can be further mined to generate useful knowledge. The convergence of connectivity, communication, interaction and knowledge enabled by human personas encompasses the ultimate function of cyberspace in modern existence.

## 2. THE INTER-CONNECTED WORLD

The theory of inter-connectivity in cyberspace holds that people, things, entities, processes are inter-connected by the collaboration of the human, social and digital world [4]. The cyber-physical-social space is an interpretation of the theory of inter-connectivity in cyberspace in which physical perceptions, cyber interactions and social correlations are inter-connected through ubiquitous virtual reality. An underlying basic assumption is a form of inter-dependence between activities in the virtual reality, activities in the physical reality and activities in the social reality which is based on a perfect integration of the cyber-physical-social space with human, social and physical interactions.

Cyberspace, artificial and virtual realities are a generalised form of digital abstractions that support inter-connectivity and interactions between cyber entities. Cyberspace is assumed to be independent of space and time constraints as its existence is a timeless virtual abstraction of the physical reality [7]. For example, the death of a person does not necessarily imply the death of associated cyber accounts. Access to cyberspace is possible via digital devices on independent networks. The cyberspace ecosystem also provides infrastructure for massive information management services, resource management and service management through public interfaces mostly running on distributed databases. This ecosystem is also supported by uniform standards and protocols to enhance the flexibility of usage, ease of access and create a self-evolving computing eco-system. Although cyberspace operates as a global domain consisting of independent sub-networks of technological infrastructure and data that potentially connects all interacting entities, the physical barriers of the real-world dimension (territory, land) are obsolete. This way, it can provide the maximum number of people with the means for increased productivity and creativity within its new information society.

The physical space refers to real-world entities that interact directly or indirectly with cyber entities. The idea of physical space saturates modern existence as a concept for understanding the world around us and how we interact with its comprising entities. Thus, it encompasses all physical entities where physical interactivity is possible, heterogeneous interfaces, physical infrastructures and interactive environment required for seamlessly browsing through cyberspace. These include; all real-world components and the physical infrastructure that support the existence of independent networks. The facets of the physical space: Location, Distance, Size and Route [8] attribute physical entities and their mappings in cyberspace. A physical object can be mapped into cyberspace using geographical attributes (IP Addresses, Longitude, Latitude) so that its location replicates where the thing exist in the physical dimension in relation to other physical entities in the same space. The amount of time required to exchange information between two physical objects is a function of the distance between the two objects, the size of data being transmitted and the channel of communication [9].

The social space is an integration of social attributes, inter-personal relationships controlled by physical people and other physical objects. A physical entity creates direct and indirect relationships with other physical entities or cyber entities, therefore, establishing an eco-system of semantic relationships that mirror human social behaviour [4]. Physical people map characteristics or attributes of their personas into the social space as cyber personas. The personas and cyber-personas characterise human activities, social events, behavioural conventions, political administrations, public services for human social participation in cyberspace. A single physical user may be associated with multiple cyber personas; however, each cyber persona can be associated with one individual. Social inter-connectedness enables the creation of online communities characterised by the active social presence, social participation, relationship creation, and collaboration and belief affiliations. These online communities sometimes mirror traditional real-world communities where members share and express similar political, cultural, social and economic ideologies. These three sub-domains converge to a cyber-physical-social ecosystem held together by four main features; interactivity [10], inter-connectivity [11], interaction [12], integration [13] and intelligence [14].

## 2.1. The 5Is Of the Multi-Dimensional Cyberspace

- *Interactivity* refers to a two-way effect created when two or more entities meet each other [10]. Entities in cyberspace can establish interactivity with other entities within their cyberspace dimension or with entities in other dimensions. For example; a web application accepts an input from a human, or a mobile phone establishes a one-touch payment process. The ease of access to cyberspace and robust infrastructure supporting distributed networks creates the platform for easy and speedy multiple interactive sessions between these cyber entities.

- *Inter-connectivity* refers to a state of being connected or the linking of two or more entities in a given space [12]. It is made up of multiple simple interactive processes happening simultaneously. Although this concept is often used when referring to the linking of two digital devices in an internet network, inter-connectivity in cyberspace broadly indicates that physical objects, people, cyber personas and social dynamics establish seamless communication or relationship links between and within each other. This implies that entities may be linked via physical connections or social relationships and interactions across the various dimensions of cyberspace. This logical infrastructure provides the foundations for transmitting information across large distributed networks and the creation of online social communities for collaboration and sharing.

- *Inter-communication* provides the structure for connecting and linking with other entities in cyberspace while interactions provide communication support between entities that lead to the seamless creation of data and information streams in cyberspace. Cross-domain data are fused by integration, which can further be analysed for actionable intelligence.

- *Integration* refers to the fusing of two or more entities in a coherent and unified manner. There may be combinations across different dimensions of cyberspace — for example, cyber-cyber, social-social, physical-physical, cyber-social, cyber-physical or social-physical. Integration in cyberspace may refer to the coming together of multiple entities in cyberspace under a uniform standard protocol or the fusion of data flowing from multiple interacting sources using a global schema and mapping.

- *Intelligence* refers to the self-enabling ability to acquire process, interpret and apply knowledge. The seamless integration of knowledge across the cyber-social-physical spaces precedes large-scale dynamic collection and processing of information, which allows for self-learning, adaptive behaviours and self-evolution of entities in cyberspace. Intelligence exists across all dimensions of cyberspace in forms of human intelligence [10], machine intelligence [15] and emotional intelligence [16].

The 5Is of entities in our physical and virtual worlds have created a new platform for dynamic information exchange via a distributed communications network. Thus, these interactions, in turn, created logical interdependence between interacting and inter-connected elements across cyberspace. Barnett [17] recognises the importance of the interdependent nature of systems, people and networks in an operating multi-dimensional space.

## 3. CHARACTERISING CYBERSPACE: A MULTI-DIMENSIONAL SPACE

Cyberspace encompasses the people interacting, the devices connected and the information flowing within it. In recent times, the role of cyberspace has gradually evolved as an enabler of day-to-day lives. Cyberspace is a global village of inter-connected entities interacting through time. An interesting term cited in [4] refers to cyberspace as a virtual and physical abstraction with all the information resources that create the interconnections among cyber entities. Consequently, cyberspace can be characterized by expanding its prior boundaries to represent links between physical and virtual entities on a time spectrum. Clark, Klimburg & Mirti [1], [2], provided a four-layer contextual model that can also be used as a structural framework for information identification in cyberspace. Arguably, these four layers can reflect the cyber hyperspace rigorously discussed by [4]. The first layer, the Physical Layer, characterises all physical devices connected in cyberspace. The second layer, the Logical Layer, characterises the logical building blocks supporting the infrastructure of cyberspace. The third layer, the information Layer characterises all information or raw data that is generated, stored transmitted and processed by cyber entities through cyberspace. The fourth layer, the People, characterizes the human element of cyberspace communicating and utilising the functionalities of cyberspace. Figure 1 is an illustration of the Multi-Dimensional Cyberspace.

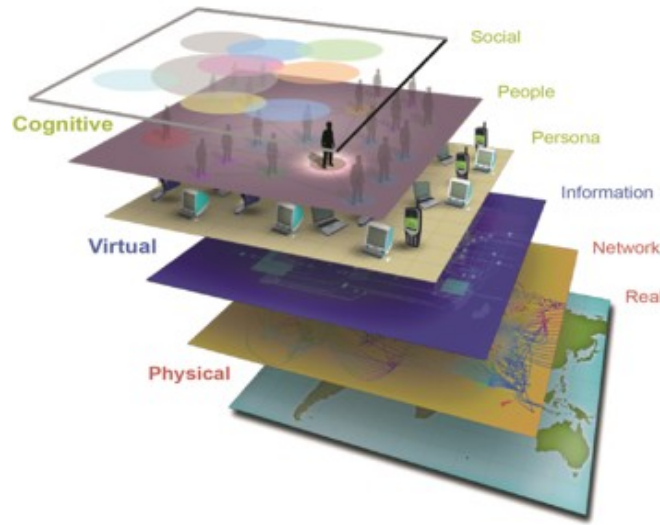

Figure 1. The Multi-Dimensional Cyberspace [18]

### 3.1. The Physical Dimension

Ning et al. [4] describe the physical dimension as a combination of the real world as perceived from a linear dimension and physical objects as they exist in cyberspace. Although Clark and Klimburg [1], [2], [19] restricted the physical dimension to all physical objects and hardware mapped into cyberspace via sensing and digitisation, some studies have advocated the real-world association to entities in cyberspace. For example, a physical object can be digitised and mapped into cyberspace with a unique IP address. However, it is also possible to map all IP addresses to a physical geo-location in the world which has attributes like weather, traffic, political and economic events, financial markets, commodity markets, economic indicators etc. Therefore, the physical dimension is not limited to include the hardware devices that support access to cyberspace but also attributes that define geographical characteristics of these entities and technical infrastructure that defines the structure of connectivity between them [19].

- *The Networks*: Sensors and activity monitors, support activities on the network layer. The networks are an embodiment of interconnected computing devices in which these devices are perceived and controlled by sensors and actuators to establish a communication link between them [19]. These include all the hardware (servers and computers), infrastructure (wired, wireless, and optical), physical connectors (wires, cables and routers) and digital devices interacting in cyberspace.

- *The Real-World*: The geographic or real-world components of cyberspace refer to the physical location of elements of the network. These are the characteristics that define the physical location of a network object. It is essential to differentiate the real-world components from activities on the economic dimension. While activities in the economic dimension relate to events that affect economic, political, social and cultural dynamics, real-world components are seen as a naturally occurring phenomenon. Therefore, the geographical components are not restricted to the physical dimension.

### 3.2. The Social Dimension

The social dimension comprises of the persona and cyber-persona components [1], [2]. The personas and cyber-personas simulate the real-world to a virtual world from social events, human activities, political administrations, behavioural conventions, public services for human social participants in cyberspace. Social inter-connectedness establishes the existence of the social dimension by enabling the creation of online communities characterised by the active social presence, social participation, relationship creation, and collaboration and belief affiliations [19].

- *The Cyber-Personas*: Within a network, components of the cyber-persona help in describing the online identities of people as they exist and interact with various aspects of cyberspace. Individual cyber-personas are closely linked to an actual person or persons incorporating some cyber characteristics, e.g., IP address, social media accounts, etc. A single individual may possess multiple cyber personas, and a single cyber-persona can have multiple users [19].

- *The Physical-Personas*: he persona refers to the offline mode of the cyber-social dimension [4]. The persona components describe characteristics of the people interacting within a network from a real-world perspective; it also helps in describing the characteristics that define them as they exist in the real world. For example, a person may have attributes like a home address, a financial service provider or even a work address [19].

### 3.3. The Economic Dimension

The economic dimension characterises cyberspace as an enabler of our socio-political-economic-cultural existence. The socio-political-economic-cultural dimension is characterised by activities attributed to human beliefs, norms, laws and co-existence. As a result, it captures the state of human experience as they exist as individuals or communities. It reflects how activities on the real-world economic and political hemisphere affect the events in cyberspace [20]. Current research has made attempts to quantify the association between socio-political-economic-cultural factors and events in cyberspace. For example, Cavusoglu, Mishra and Raghunathan [21] studies the effects of various types of cyber-attack announcements related to stock market prices and Gandhi et al. [22] explores the effects of economic, social, political and cultural indicators on the occurrence of cyber-attacks. Although there has been a minimal focus on the direct impact of stock and commodity market price fluctuations on the event of cyber-attacks, there is substantial evidence supporting their influence across multiple layers in cyberspace [19], [20], [23], [24]. Recent activities in cyberspace also prove how economic, political and cultural dimensions of society lead to cyber-incidents [25], [26].

- *The Broadcasts*: The high volume and velocity of content being generated in real-time of real-world incidents support the existence of this layer. The information in this layer also describes events in the real world. A unique characteristic of this layer is its real-timeliness and newness of information generated [19]. These are facts or 'knowns' and include information from certified media sources. While some may argue that most social media platforms such as Facebook, Twitter, etc., can also act as news dissemination actors [27], the news is seen to be biased in such aspects by the uncertainty of the information sources.

- ***The Markets***: This layer describes the effects of trade, money and even finance on modern existence and how they act as triggers to cyber defence events [19]. Hence, it is critically important to identify what market forces influence cyber defence incidents. Markets within this context refer to all financial, trade and money context of the real world as reflected in cyberspace. For example, stock prices, money demand, crude-oil prices, money supply, inflation rates, fluctuations in exchange rates, deflation rates may have impacts on economically or politically motivated cyber-attacks direct or indirectly [19]. Moreover, Bollen [23] proves that elements in cyberspace can influence stock market prices as well as market valuations being influenced by denial-of-service attack announcements on the market valuation of firms [19], [21]. Similarly, Bronk and Tikk-Ringas [28] explain how events in the oil and gas sector affect the propagation of cyber incidents.

## 4. DATA STREAMS IN CYBERSPACE

Data is continually generated in all dimensions of cyberspace through interactions, activities, relationship formation and events. Data in the social space exist as a series of interconnected events that captures the real-world personas as well as the cyber personas of online users. Interactions are achieved through effortless ease of access to cyberspace while interconnections are established in local and global domains during which real-time data is collected, sanitised and stored for future intelligence analysis. Activities in the physical, social, real-world or cyber dimensions are initiated by entities within contextual domains creating historical maps of event-based correlations [29]. Relationships are formed based on common social, cultural, political, economic or technological beliefs creating associations between entities. Information collection, processing and storage are possible through high-performance digital devices that self-enable knowledge creation and dissemination. The social space could be described as a complex interactive structure made up of organisations, individuals, and entities connected by one or more specific types of inter-dependencies such as friendships, interests, sexual relationships, political, likes, dislikes, economic and religious affiliations, relationships of beliefs, kinships, etc. [19]. Widely accessible and open source social technologies are increasingly being used by humans to share thoughts, perceptions, opinions and beliefs about real-world events in real-time. Social media platforms such as Twitter, Facebook, YouTube produce streams of information capable of creating proactive intelligence. Cyberspace creates a platform for individuals to operate multiple personalities with which they can share true opinions freely, beliefs and associations (Personas and Cyber-Personas) [17]. A typical example is the formation of online hacker communities', e.g. hacktivists, cyber soldiers as a platform to plan, coordinate and execute cyber-attacks.

The physical space consists of real-world dimensions that are mapped into cyberspace. These could be the digital devices through which cyberspace is accessed or real-world perceptions of geographically distributed cyber entities. Data collection is enabled by using sensors, actuators and context-aware networks [6]. Methods for collection, storage and knowledge extraction on Human-Machine interaction data being generated in real-time for example; network traffic flow [30] can be used to monitor network access [31], spot malicious behaviours such as denial of service attacks [32], build models for automating security protocols and understand network behaviour [33]. Typically, software applications hosted on the world wide web could be configured to collect and store interaction data and user data [34]. End-user behaviour could be inferred using event logs for evidence gathering [35] and anomaly detection techniques [36]. Activity log data collected from mobile devices or activity devices could be used in health intelligence. Finally, data in the real-world dimension such as GPS tracking data, weather data and transport traffic data could be integrated for informed intelligence to trace events such as cyber terrorism or cyber activism [37]. Due to the sensitivity of information generated in the physical space and the risk involved with activities around it, data on this domain is not open or publicly accessible. This major challenge could hinder efforts at developing generalised techniques for knowledge extraction [38].

## 5. CONCLUSION

The decentralisation of networks also creates an open market for ideas and beliefs that are often shaped by immediate surrounding events. The complexity of the relationships between real-world events and cyber events provides a new dimension for analysis. Historical and current correlations between social, political, economic and cultural events [22] could provide actionable intelligence for cyberspace operations. Similarly, in a contest of competing narratives, the concept of 'news' becomes relative and what is accepted as truth could be considered important when dealing with global events. There is sufficient evidence to prove that useful intelligence can be gathered from a systematic analysis of unstructured open source information, especially information from the media [39]. For example, studying past social events in news media and publications to explore factors in the social and cultural dimensions that act as antecedents to cyber incidents [20].